# Contextual and Structural Representations of Market-mediated Economic Value

**Bradly Alicea**
**Orthogonal Research, Champaign, IL 61821**
bradly.alicea@outlook.com

**ABSTRACT**

How do we assign value to economic transactions? To answer this question, we must consider whether the value of objects is inherent, is a product of social interaction, or involves other mechanisms. Economic theory predicts that there is an optimal price for any market transaction, and can be observed during auctions or other bidding processes. However, there are also social, cultural, and cognitive components to the assignation of value, which can be observed in both human and non-human Primate societies. While behaviors related to these factors are embedded in market interactions, they also involve a biological substrate for the assignation of value (valuation). To synthesize this diversity of perspectives, we will propose that the process of valuation can be modeled computationally and conceived of as a set of interrelated cultural evolutionary, cognitive, and neural processes. To do this, contextual geometric structures (CGS) will be placed in an agent-based context (minimal and compositional markets). Objects in the form of computational propositions can be acquired and exchanged, which will determine the value of both singletons and linked propositions. Expected results of this model will be evaluated in terms of their contribution to understanding human economic phenomena. The paper will focus on computational representations and how they correspond to real-world concepts. The implications for evolutionary economics and our contemporary understanding of valuation and market dynamics will also be discussed.

## Introduction

What is your alien artifact worth? This seemingly strange question has a less ambiguous motivation centered upon how the economic value of objects and ideas gets determined. Are there a set of intrinsic properties and mechanisms that determine value, or is valuation purely a product of human interactions and cultural construction? And if the answer lies between these two viewpoints, then how can it be represented? To better understand this, a somewhat false dichotomy will be introduced. This will motivate an evolutionary approach that bridges both of these perspectives.

## Origins of Economic Value

To computationally model the origins of economic value, we must understand valuation at the levels of markets, behavior, and neural mechanism. On one hand, we have the efficient markets hypothesis (EMH). In the "strong" formulation of EMH, prices fluctuate according to a random walk, and thus cannot directly predict value [1]. Yet information about prices (and ultimately value) is contained in the market's structure, particularly when the market is in an equilibrium state. Other formulations of the EMH (e.g. "weak" EMH) suggest that there are fundamental valuation metrics that may allow for partial prediction with respect to market prices [2]. Alternately, we have the idea that valuation is a product of cultural construction and interactivity. From a cross-cultural and historical perspective [3], it is apparent that value is a relative property that is dependent on systems of beliefs and morality [4, 5]. Contrary to the EMH, behavioral and cognitive factors play a more prominent role in determining how individuals interact with markets. Rather than discovering value from the intrinsic properties of a market's collective dynamics, the interaction of thinking agents and the multitude of meanings (polysemy) each agent attaches to an object generates a unique valuation scheme with respect to context.

Aside from this argument, biological and cross-species comparisons reveal that valuation seems to be both a universal feature of all human societies and a conserved feature of primates. A meta-analysis [6] suggests that grooming and agonistic support may have evolved as part of a low-cost reciprocal altruism system. While this serves as a risk-management mechanism and buffer against the negative effects of competition [7], reciprocal grooming may also function as a *de facto* currency of exchange. Likewise, the human brain network for valuation has two components. These subnetworks govern the evaluation of stimuli and events in the world and the ability to make judgements on these observations, respectively. The risk and reward subnetwork involves projections from dopaminergic neurons in the brainstem (e.g. substantia nigra) to the ventral striatum (e.g. nucleus accumbens). The value assignment subnetwork involves mediating projections from the brainstem to the prefrontal and orbitofrontal cortex. This has been related to behavior by [9], who demonstrate that introducing performance-based rewards on tasks reduces performance and shows a decrease in dopamine and brain activity associated with reward processing. These brain centers and signals provide a neural architecture for higher-level cognitive processes such as decision-making and natural classification.

### An Alternative Approach and Synthesis

To better understand the evolutionary emergence and nature of economic value, the concept of cultural markets will be introduced. Cultural markets have the potential to resolve theoretical tensions between proponents of self-regulating markets and the cultural construction of economic value (see Figure 1). In terms of structure, cultural markets are similar to the concept of biological markets [10] in that interactions between individuals are contingent upon both diversity and evolution. Implicitly, the neural substrate of Primate valuation is assumed to drive interactions between individuals. Explicitly, cultural context and market interactions are an important driving force behind the relative value of objects and exchanges.

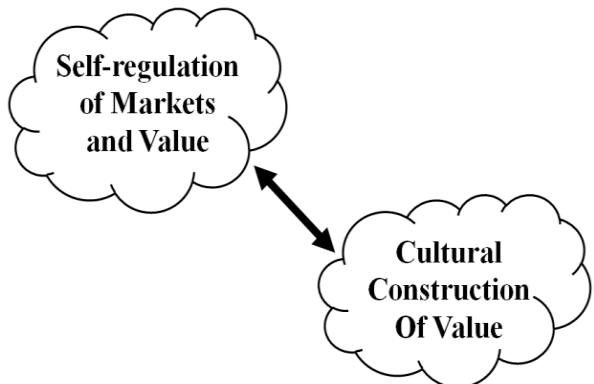

**Figure 1.** Graphical representation showing the competing ideas of self-regulating markets and cultural construction of value without markets.

**Value as Interactive Cultural Behavior**

We can use an agent-based approach which enables cultural (e.g. culturally-specific) behaviors to model cultural markets and the evolution of value. Contextual Geometric Structures (CGSs) can be used to represent the relationship between objects of the same type, or different object types. CGS kernels are soft classifiers introduced in [11] that capture cultural behavior with respect to cognition, neural processing, and collective behavior. This model bridges the implicit and explicit aspect of valuation by modeling the neural substrate as a set of conditioned perceptual classifiers that incorporate an artificial genome.

The assumed connection between fluctuating information, perceptual classification, and a basis for value requires the incorporation of new information. The architecture of a scaled CGS kernel is shown in Figure 2. As with the standard CGS approach [11], the *n*-tuple representation embedded in an *n*-dimensional space. These dimensions are derived from relevant sensory processing and cultural bases. Using the CGS soft classifier is useful for representing the effects of culturally-mediated conceptual blending on perceptual classification and ultimately the assignation of value. This allows for diversity of valuation criteria across populations, in addition to an adaptive and abductive decision-making framework. Yet the Primate neural substrate demonstrates that valuation does not rely solely upon a single act of natural classification. Rather, a complete assessment of value involves many centers which include classification, decision-making, and references to attention and memory. Two advances to the conventional CGS approach involve kernel scaling and the addition of a genotypic representation. Kernel scaling is necessary for the incorporation of novel, previously unobserved information. A tuning parameter ($\alpha$) governs the extension of the kernel along each dimension. The genotypic representation consists of three sets of genes: genes that define the dimensions and anchors, genes involved in the cryptographic hash, and genes that define an agent's inherent mental flexibility. This type of representation is potentially more compact than a very large connectionist network, but retains the generative capacity of such networks. More details on the scaled kernel design can be found in Methods, Section 1.

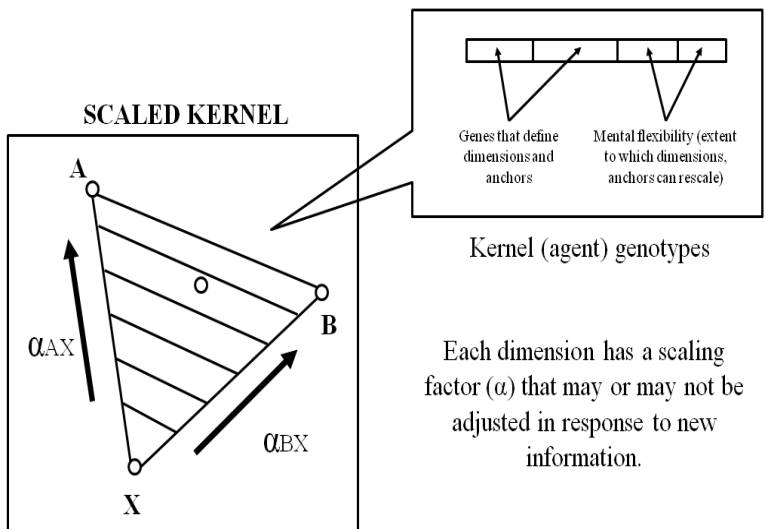

**Figure 2.** Example of a scaled kernel, adjusted by a given kernel's response to new information. INSET: 1) genes that define the dimensions and anchors, 2) genes involved in the cryptographic hash, 3) genes that define an agent's inherent mental flexibility.

**Transactional Representations**

This approach requires additional representations that define a medium of exchange and a transaction space. Our medium of exchange involves the use of a currency. As a cross-cultural phenomenon, currencies provide a means to standardize value across transactions, and serve as a placeholder for economic transactions. While the form of currency varies across social context (e.g. paper money, shells, houses, bits), its function as a placeholder is generally the same. We can use a placeholder similar to digital crypto-currencies for purposes of assigning value to transactions in parallel. Besides allowing us to quantify economic transactions, the crypto-currency approach [12] allows us to assign a unique identifier to each transaction. The Bitcoin model uses a cryptographic hash function [13] to enable secure transactions. We can use a similar mechanism (Methods, Section 2) to distinguish bidding behavior from complete transactions.

The concept of biological markets [14] can be used to define a transaction space for a population of agents. Scarcity is used as a criterion to assess the need for exchange, make selective transactions, and to minimize cheating. However, information concerning the objects and their relative worth also plays a role. Let us consider how the brain produces a judgment of value. When a transaction is made, an object is offered by the seller at a given price. That initial price is either accepted or rejected by the buyer. This leaves the flexibility of value for a given object open to one's perceptual classification of the natural world. The synthetic biological markets introduced here are represented by a dual mechanism: the minimal market and the prediction (or compositional) market. To understand the role of each mechanism, it is important to point out that CGSs enable non-transitive relationships between all possible combinations of agents and placeholders, agents and objects, and placeholders and agents. Figures 3 and 4 demonstrate how these transactions are structured. Figure 3 illustrates how the cryptographic hash concept can be applied to determining economic value. In a cryptographic hash, a transaction between agents consists of two parts: an object and a hash function. In Figure 4, the object is presented to the agent. However, the agent can only purchase the object if the placeholder allows them to. In this case, the placeholder is a hash function based on an internal threshold based on consistency with their CGS classification (see Methods, Section 3 for more information).

**Types of Markets**

The first mechanism used to approximate the market for objects utilizes the *minimal market* approach (Figure 4A) to model pairwise exchanges of single propositions between agents. When embedded in a minimal market, agent-agent interactions create a generative search space, in which outcomes that have not been predetermined can be explored. In this way, there is no direct correspondence between fluctuations in the value of placeholders (e.g. money) versus the value of objects. However, there is a more general conceptual relationship that is established between objects and the placeholder. This allows for agents to "print money" and generate wild speculation as required by context. The second mechanism, used to approximate the market for placeholder value, utilizes the *compositional market* approach (Figure 4B) to model top-down mediated auctions that set value for linked propositions. Compositional markets (as opposed to minimal markets) involve large-scale (one-to-all) transactions. In Hanson [15] and Dudik et.al [16], such markets are centralized sources for predicting the value of a certain set of political scenarios. This allows agents to collectively set the relative value of objects, which ultimately scales up to a coherent economic system.

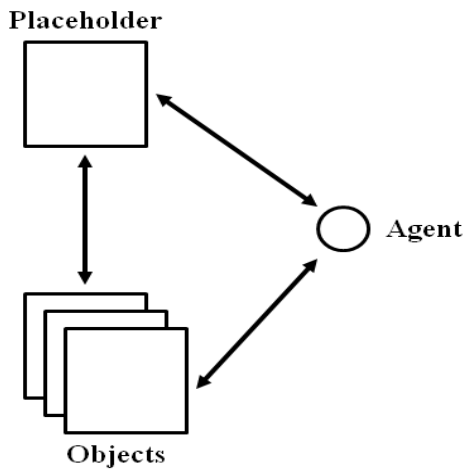

**Figure 3.** Relationship between an agent (buyer or seller), a placeholder (currency), and a type of object.

Agents can also be networked so that information can be exchanged between them. When agents are actively engaged in transactional behavior, an arc is established. In previous version of the CGS model, agents were allowed to move around the field in a manner determined by the environment. In this version, an agent's mobility is determined by its previous interactions. Clustered agents represent those with a common initial classificatory scheme. However, with the addition of kernel scaling, nearby agents can be disconnected. In a like manner, initially dissimilar agents (e.g. those in a different cluster) can be connected, producing heterogeneous markets.

## Predicted Results

**Value Fluctuation**
In cases where either a single proposition or linked set of propositions is hard to classify, the result is a high degree of fluctuation in value. This result will typically be undergirded by objects that are popular but not well understood in terms of their composition or phenomenology. Examples of this from the real world may include pricing bubbles, new technology hype, and even the value of popular entertainers and political figures. We can analytically demonstrate how these fluctuations might emerge using exemplars embedded in the compositional and minimal market configurations. These examples will elucidate how value can fluctuate across individuals making purchases from single sources as well as between individual agents. This will hopefully suggest ways in which the concept of value fluctuations (as opposed to *de novo* value construction) is multidimensional.

In compositional markets, a series of proposition become functionally linked together. This series includes a full set (e.g. A-B-C) in addition to singletons that form a transitive relationship (e.g. A, B, and C). A fluctuation in this experiment will be revealed by large differences in the value of the full set (A-B-C) relative to all subsets. A secondary criterion for fluctuation can be revealed if value is divergent from the rules of logical transitivity. For example, if B is worth more than A, and C is worth than B, then A will be worth more than C. When these propositions are functionally linked together, some non-transitivity will necessarily result, as the gain in value for each combination is not always linear. However, it is when there is a large (e.g. many-fold) gain or loss that a fluctuation in value occurs.

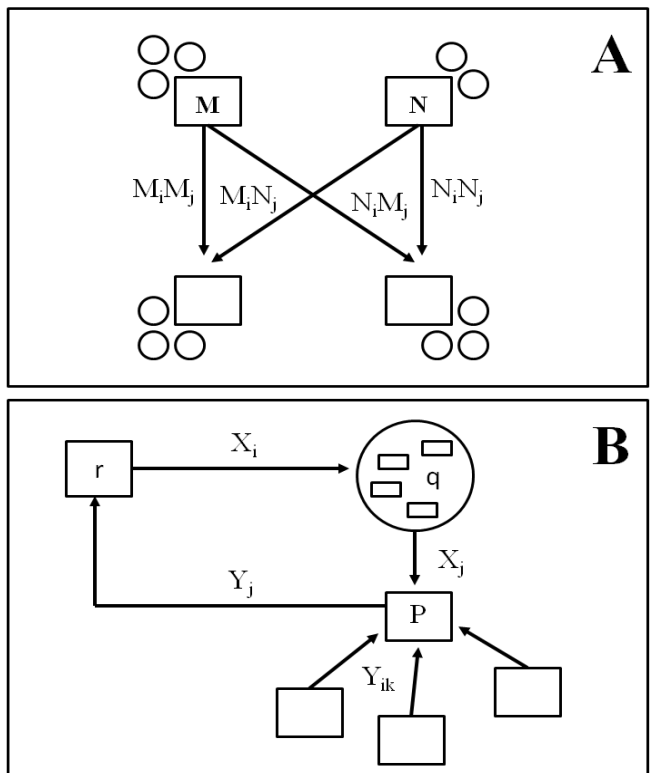

**Figure 4.** Example of synthetic market structure as represented by two components: minimal (A) and compositional (B) markets. See Methods, Section 3 for notational definitions.

We can also show that in minimal markets (Figure 6), the highly asymmetric exchange of singletons and subsets of A-B-C also provide a window into possible fluctuations in value due to conceptual (e.g. cultural) differences. This is particularly likely when the pairwise exchange is made between agents with different CGS schemes. For example, if the item being exchanged is something the buyer has prior experience with, then the value is likely to be similar to the object mean [17]. A different situation holds true for "missing piece" singletons, which can be applied to an agent's complement of propositions to hold something of much greater value than would otherwise be the case.

**Large-scale Gains and Losses in Value**
Given the run-ups in value (sometimes called bubbles) in real-world markets ranging from tulip bulbs to housing, we can model these run-ups not as anomalous events, but as fundamental features of a market. The origin of such run-ups can be reduced to pairwise transactions shown in Figures 7 and 8. When replicated across the network topology, such run-ups can result in bona-fide bubbles. Figure 7 outlines the pairwise underpinnings of these types of transactions, while Figure 8 demonstrates how gains in value can cascade across a network of agents over the course of 3-5 transactions. Once a transaction is established between agent A and agent B, directionality is established which sends a signal of net gains (and net losses) to the rest of the network for this particular transaction. In this example, agent A makes a transaction with agent B. In the process, agent A sees a net gain of 20% (in arbitrary units). While this might be an extremely local (e.g. in the context of agent A and agent B) phenomenon, it can nonetheless be copied by other transacting agents.

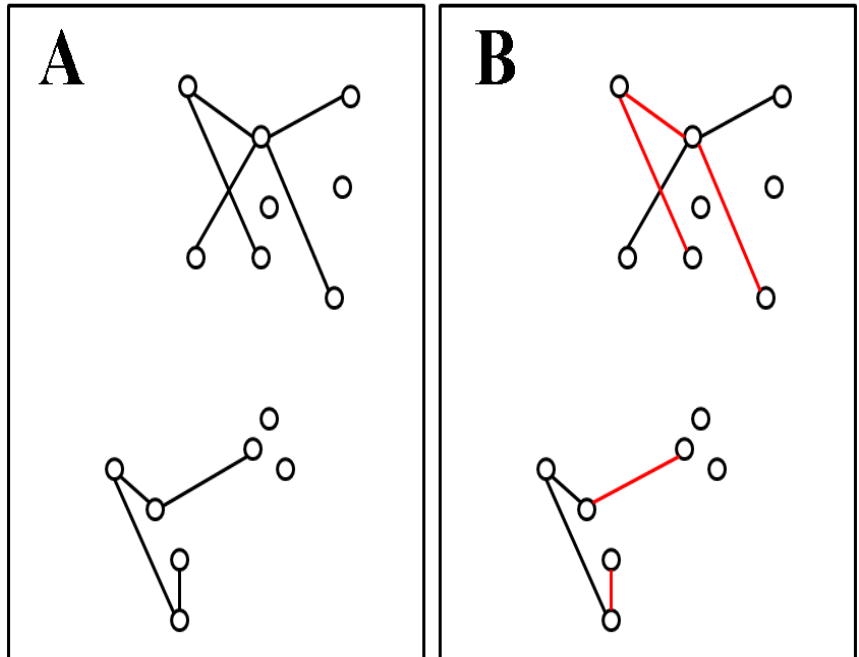

**Figure 5.** An example of connectivity between agents at single points in time. A: connections between agents that are currently engaged in a transaction (black arcs). B: connections between agents that are currently engaged in a transaction (black arcs) and transactions between agents with a common scaling factor (red arcs).

By using imitation rather than cultural and neural classification as a decision-making heuristic, a cascade of asymmetrical transactions can result. Methods, Section 4 provides more information as to how this imitation mechanism works in the model. More details about the observation criterion in the service of a selection function can be found in Methods, Section 5. One thing that stands out in Figure 8 is that a valuation bubble can be sustained through feedbacks. This behavior resembles an overloaded power grid, in which runaway feedbacks contribute to catastrophic failure. In this case, if an asymmetrical exchange takes place somewhere in the network, first-order connections can observe and replicate this without using a formal criterion. This is consistent with the multiple mechanisms for valuation in the brain.

**Value Inequality and Social Evolution**
Now we turn from the emergence of valuation at the individual scale to interactions at the societal scale. This involves not only pairwise interactions in a network, but also first-order observations (or observations of direct connections in a network). This first-order observational capacity serves as a bridge between the pairwise and compositional market configurations. This bridge is an essential feature of a valuation model, particularly for capturing herding and other mob-like group behaviors. In general, herding and other group behaviors provide a mechanism for understanding valuation bubbles.

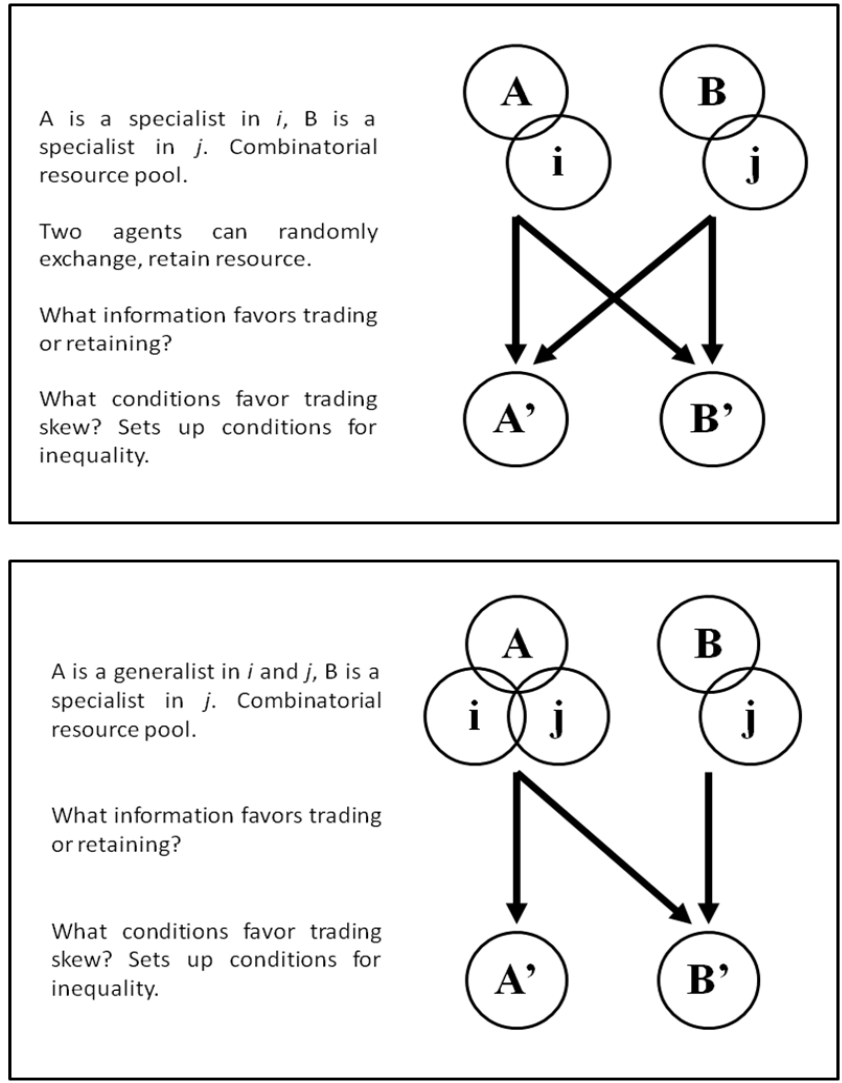

**Figure 6.** The concept of a minimal market, using a pairwise example (A, B). **TOP:** symmetrical trading between two specialist agents. **BOTTOM:** asymmetrical trading between two generalist agents.

Demonstrating how this model can be applied to networked microeconomic scenarios allows us to move towards a type of meso-economics. Instead of generating theoretical models that assume equilibrium, the quasi-evolutionary approach taken here provides a series of plausible scenarios for social evolution. It is the social and cognitive plausibility of economic interactions with respect to valuation that can potentially address the inconsistencies of existing perspectives.

As a representation of the parallel economic transactions that underlie social evolution, this model presented may also capture both neurocognitive and social aspects of valuation over time. It is the economic evolutionary capacity which is potentially the most promising aspect of this approach. This model allows us to capture both uncertainty in markets and generating a range of economic behaviors. In addition, the configuration of this model is meant to capture observed cultural co-evolution between market dynamics and social cognitive capacity.

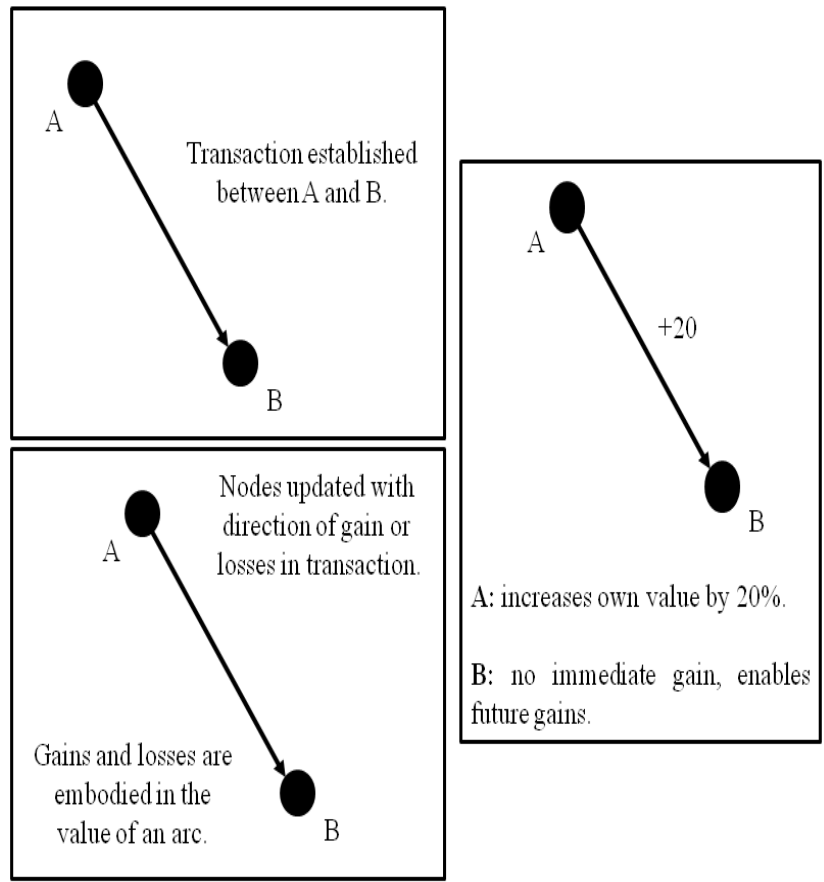

**Figure 7.** Summary of network approach using a pairwise (A, B) example. UPPER LEFT: A transaction is established between A and B. LOWER LEFT: definition of net losses and gains. RIGHT: example of net gain (20%) or agent A during a transaction.

**EMH-like vs. Polysemy**

While this is not a predictive model, we can nonetheless use our potential measures to determine whether or not the values generated for an object were done so based on a random (efficient markets) mechanism or a polysemic (culturally-specific) mechanism. Figure 9 shows a hypothetical bivariate relationship where the range of acceptable values ($Y_{ik}$) predicts the soft classification value ($A_x$). In this case, a 2-tuple CGS would be the dependent variable (y-axis). However, for more complex CGSs, multiple dependent variables could be used. In these cases, the nature of value can be characterized in a multi-dimensional space.

In the bivariate case (Figure 9), there are three regimes: strong polysemy, weak polysemy, and EMH-like. They are defined not only by their extent in bivariate space, but their spread along each axis. For example, when the data show significant variance along the x-axis, the process is randomly walking in value space, which is indicative of an EMH-like valuation process. By contrast, when the data show significant variance along the y-axis, the process is randomly walking in meaning space, which is indicative of a polysemic valuation process. Strong and weak polysemy can then be defined as the orientation of outliers along each axis. For strong polysemy, there is fewer outliers along the x-axis than along the y-axis, while in cases of weak polysemy, the opposite result is expected to hold.

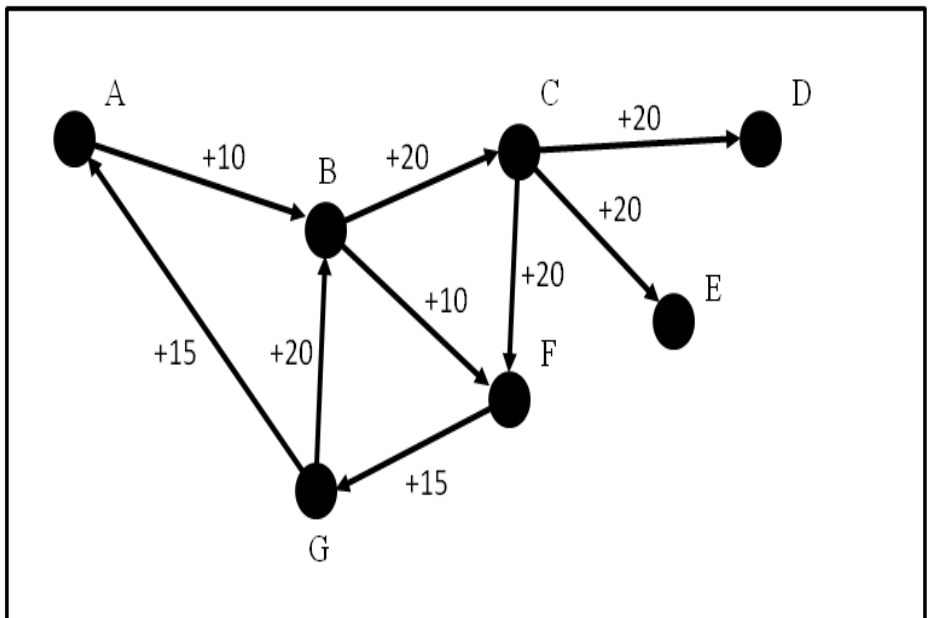

**Figure 8.** A series of transactions on a seven-node network. Universal gains in value for pairwise transactions reveal an emerging bubble.

## Conclusions

While the idea of an economic bubble is hard-to-define, this work can contribute to a more rigorous definition. While bubbles are generally thought of as rises in asset prices relative to value, they might be predicted by detecting deviations from fundamental value [18]. As to defining exactly what goes into determining fundamental value, this work can contribute to understanding what information, cognitive biases, and social relationships go into the valuation of objects and transactions. By employing the CGS approach, we can provide a link between bio-inspired models such as neural modeling fields [19], quasi-folk classification systems, and conventional models of trading and economic exchange.

This approach is consistent with cognitive neuroscientific [20] and agent-based [21] models of valuation and evolutionary economics. At the individual level, there are multiple mechanisms involved in valuation [22]. At both the individual and population levels, uncertainty also plays a role in valuation [23]. This is particularly true of large fluctuations and degrees of variation in value. Yet whether these values come from the market itself or from cultural context can be tested experimentally. For example, when value converges across groups with a diversity of kernel compositions, then this is strong evidence that value comes from the market itself. By contrast, when value is highly divergent between groups, then this is strong evidence that value comes from cultural context. One outcome yet to be shown is that market success based on popularity or immediate payoff (e.g. the traditional supply-and-demand model) is a suboptimal outcome. Such success is based on an inherent overvaluation due to the lack of searching the market space for a global equilibrium.

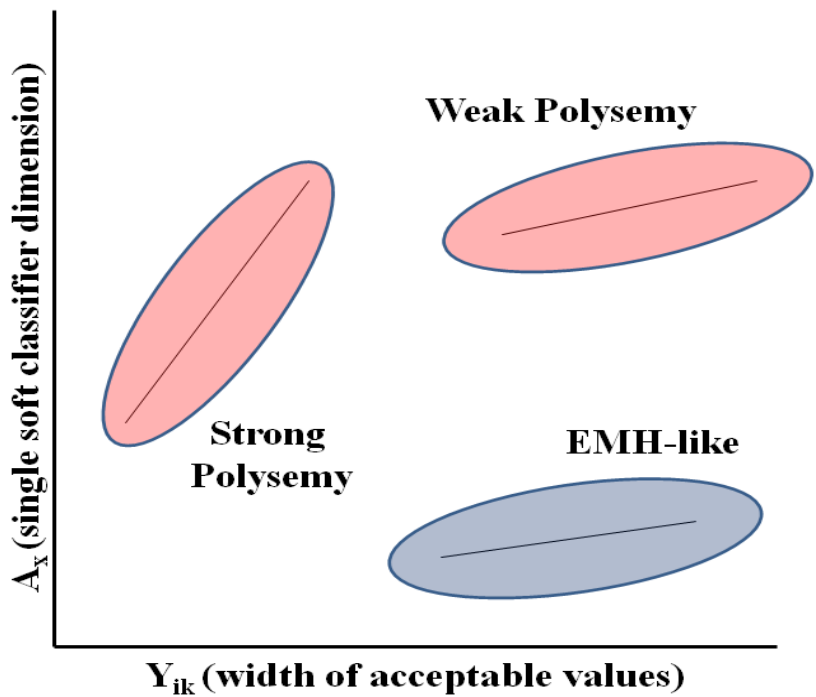

**Figure 9.** Expected bivariate relationship demonstrating how the range of acceptable values should be related to the soft classifier values. This graph is also a representation of value space ($Y_{ik}$) vs. meaning space ($A_x$).

One advantage of this approach is the flexibility to model not only tangible goods and financial forms of capital, but social and political capital as well. In these cases, the objects are eliminated and the object of valuation using the placeholder functions becomes the mental representation of the agent itself. In a typical economic model, non-tangible forms of exchange and valuation would be a nonsensical set of arguments. Non-tangible ideas can be better valued using a model such as the one described in this paper.

Another future direction for this work involves distinguishing between markets rich in information and markets which are information-poor. Recall that conventional EMH theory predicts that all information is contained in the market. However, the twin dicta "the wisdom of crowds, the madness of crowds" suggests that as collective behavioral structures, economic markets can be both.

**Methods**

**Section 1: formal definition of a market**
The notation for Figure 3 is as follows. In A, pairwise exchanges between agents *M, N* take place in a minimal market. An interaction is either made to the same agent over time (e.g. $M_i, M_j$), or between agents over time (e.g. $N_i, M_j$). Two semi-independent exchanges are made: objects (*i*) and value (*j*). In B, an operator (*r*) functionally links sets of propositions together ($X_i$) into a linked ensemble (*q*) and offers them at a pre-determined price (*X<sub>j</sub>*). The offer is evaluated by traders (agents in the market) who decide to accept the offer based on a decision-making criterion (*Y<sub>ik</sub>*). *Y<sub>ik</sub>* is defined as

$$Y_{ik} = \begin{cases} \text{if } \sup(k) \leq X_j \geq \inf(k), \text{then } 1 \\ \phantom{\text{if }} \text{else } 0 \end{cases}$$  [1]

where *k* is the threshold parameter *k* =*T/j*, and *T* is a range of acceptable values. If the price (value of ensemble *q*) is too high or too low, then the set of linked propositions and their collective value can be readjusted through a feedback mechanism (*Yj*).

**Section 2: modified cryptographic hash**
Our modified cryptographic hash can be defined as a cypher which encodes an encountered numeric value and maps it to a string encoded in the agent's genome. For example, a numeric value (e.g. 545) is converted to an alphabetical string (e.g. xyy). The encoded string is then mapped to a longer (endogenous) alphabetical string in the agent's genomic representation (e.g. xxyyzz). A match in the correct order serves as a key that the presented value is within the range of acceptable values for that individual.

As the endogenous alphabetic string is part of a genotypic representation, mutations (induced either randomly or deterministically) may be allowed to occur. This allows for the possible range of acceptable values to change. This frame-shifting property may also allow for the shorter (converted) string to be culture-specific [24], or for structured variation to be exhibited across the population.

**Section 3: modified CGS details**
The tuning parameter ($\alpha$) can be split into n factors, where n is the number of dimensions extending away from the direction of the scaling expansion. In Figure 4, there are two tuning factors: $\alpha_{Ax}$ and $\alpha_{Bx}$. Due to the nature of the soft classifier, which is classifies incoming stimuli relative to previously-observed stimuli, an extension of the model is necessary for dealing with unusual and rare events. The factors of parameter $\alpha$ are adjusted on the basis of the genomic representation for mental flexibility and the initial definition of dimensions and anchors.

The inset in Figure 4 shows that the rescaling is not to the new stimuli themselves, but acts to expand the classifier to approximate its suitable location in conceptual space. There is a degree of geometric constraint to the new boundaries of this space which allows the kernel to expand beyond the scope of the new stimuli. Rather than expanding to a leading edge of newly-encountered stimuli, the new stimuli are instead framed in a new context.

**Section 4: imitation as decision-making heuristic**
In cases where objects are not well-known or defined, simple imitative behaviors can be used as a decision-making heuristic. Imitation is done by making a copy of the endogenous alphabetic string described in Section 2. This is a cheap adaptation in the sense that modifications to the CGS kernel are not required. In this sense, imitation overrides cognitive classification and resembles an emotional response. However, imitation is also imprecise in that such decisions do not take advantage of generative context or reasoning mechanisms.

**Section 5: Observation (selection) criteria**

Acquisition of new propositions (including linked propositions) in the pairwise model is determined by the observation criterion. The observation criterion for agent *x* is based in part on the existing share of singleton and linked propositions for agent *x*. Agent *x* also incorporates the behavior and proposition accumulation of first-order neighbors (e.g. agent *y*). The criterion (or decision-making heuristic) *c* defines the minimal utility of acquiring new propositions. This can be defined mathematically as

$$c_x = ( \sum \frac{K_x}{K_1}, \frac{K_x}{K_2}, ....\frac{K_x}{K_n}) * (\frac{Sm}{Sc})$$

[2]

where $K_x/K_n$ is a pairwise comparison between agent *x* ($K_x$) and its first-order neighbors (1, 2,.....n) including agent *y*, Sm is the symbolic premium, and Sc is the scarcity premium. As a departure from conventional models of utility maximization, the *c* parameter includes premiums for symbolic considerations that are treated on par with the scarcity premium. As such, both the symbolic and scarcity premiums are sampled through observations of the local interaction network.

**Acknowledgements**

I would like to thank the global financial crisis of 2008 for drawing my attention to the problems and readings that ultimately lead to this paper. I would also like to thank the BEACON Center at Michigan State University for hosting a talk that helped me develop these ideas. This work is self-funded.